\documentstyle[12pt]{article}
\begin{document}

\baselineskip=22pt
\vspace{8cm}
\begin{center}
{\Large\bf Quantum Anti-de Sitter Space}

\bigskip
\bigskip

\bigskip

Zhe Chang

\medskip
{\em
Max-Planck-Institut f\"{u}r Physik\\
Werner-Heisenberg-Institut\\
F\"{o}hringer Ring 6, D-80805 M\"{u}nchen, Germany}

\medskip
and
\medskip

{\em
Institute of High Energy Physics, Academia Sinica\\
P.O.Box 918(4), 100039 Beijing, China}

\end{center}

\bigskip\bigskip

\bigskip

\centerline{\bf Abstract}
\medskip
The quantum Anti-de Sitter (AdS) group and quantum AdS space is discussed.
Ways of getting the quantum AdS group from real forms of quantum orthogonal
group are presented. Differential calculus on the quantum AdS space are
also introduced. In particular, reality of differential calculus are given.
We set up explicit relationships between quantum group
and quantum algebra, which can be refereed as the quantum counterpart of the
classical exponential. By this way, quantum AdS algebra is deduced from
conjugation on the quantum AdS group.

\bigskip\bigskip

\newpage

\section{Introduction}

The classical Anti-de Sitter (AdS) space can be defined as the
four-dimensional hyperboloid
$$\eta_{ab}x^ax^b=-\frac{1}{a^2}$$
in $R^5$ with Cartesian coordinates $x^a$, where $\eta_{ab}={\rm diag}(-1,~1,~
1,~1,~-1)$. In polar coordinates $y^\mu=(t,~r,~\theta,~\phi)$ the line element
may be written
$$g_{\mu\nu}dy^\mu dy^\nu=-(1+a^2r^2)dt^2+(1+a^2r^2)^{-1}dr^2+r^2(d\theta^2+
\sin^2\theta d\phi^2)~.$$
The symmetry group (isometry
group) of the classical AdS space is $SO(3,2)$, which plays the role of the
Poincar\'{e} group on Minkowski space. In fact, the Poincar\'{e} group can
be obtained from $SO(3,2)$ by a contraction.
The AdS space has rather peculiar features. Its time-like geodesics are finite
and closed. The time translations are a $U(1)$ subgroup of $SO(3,2)$. The
space-like geodesics are unbounded. Furthermore the casual structure is
somewhat complicated.

The AdS group, AdS space and quantum field theory based on it
has been an interested topic for a long time\cite{01}.
In the early 80's, there was great interest in four-dimensional $N$-extended
supergravities for which the global $SO(N)$ is promoted to a gauge
symmetry\cite{02}.
In these theories the underlying supersymmetry algebra is no longer Poincar\'{e}
but rather AdS.
An important ingredient in these developments was that the AdS$\times S^7$
geometry was not fed in by hand but resulted from a spontaneous
compactification, {\it i.e.}, the vacuum state was obtained by finding a stable
solution of the higher-dimensional field equations.
There has recently been a revival of interest in AdS space brought about by
the conjectured duality between physics in the bulk of AdS and a conformal
field theory on the boundary\cite{03}. This is one of the most important
progresses in the so-called non-perturbative superstring theory. Many new
results have been obtained by making use this conjecture\cite{04}\cite{05}.
Here the AdS geometry plays a central role.
It might seem very strange that quantum theories in different space-time
dimensions could be equivalent. This possibility is related to the fact
that the theory in the large dimension is (among other things) a quantum
theory of gravity. For such theories the concept of holography has been
introduced as a generic property. The Maldacena conjecture is an example
of realization of the holography.

According to general relativity, gravity is nothing but space-time geometry.
The poor understanding of physics at very short distances indicates that
the small scale structure of space-time might not be adequately described by
classical continuum geometry.
It is long been suspected that the noncommutivity of space-time
might be a realistic picture of how space-time behaves near
the Planck scale where strong quantum fluctuations of gravity may make points
in space fuzzy. As it is well known, superstring theory is now the only
possible candidate for quantum theory of gravity\cite{06}. The recent efforts to unify
by non-perturbative dualities all the known five perturbative superstring theory is
referred as M theory\cite{07}. Many physicists believe that the M theory is a fundamental
quantum theory in eleven dimensional space-time. The BFSS matrix
model\cite{08} is proposed for the microscopic description of M theory in discrete
light-cone quantization. The basic block in the matrix model is a set of $N$ partons, called $D_0$-branes,
on which strings can end. A novel feature of the M(atrix) theory is that the
nine transverse coordinates of the $D_0$-branes are promoted into $N\times N$
Hermitian matrices. Then the noncommutative geometry becomes a natural way
to deal with spaces whose coordinates are noncommutative.

Because the noncommutative geometry description is a strong candidate of
realistic pictures of space-time behaviors at Planck scale and the Madacena
conjecture a concrete realization of holography --the generic feature of
quantum theory of gravity, it is nature to assume that the proper geometry
for the Madacena conjecture be the noncommutative AdS space.

In this paper, we discuss the quantum AdS space and quantum AdS group by
investigating real forms of quantum orthogonal group and quantum orthogonal
space. Differential calculus on the quantum AdS space are presented. In
particular, we set up relationships between quantum group and quantum
algebra, which may be referred as the quantum counterpart of the classical
exponential. It is the relationship which deduces the quantum AdS algebra naturally
from the conjugation operation on the quantum AdS group. The presentations of
the quantum AdS algebra are foundations for constructing quantum field theory
on the quantum AdS space.

This paper is organized as follows. In Section 2, we present some preliminary
knowledge about quantum group theory and differential calculus on quantum orthogonal
space. A realization of the quantum algebra $U_q(so(5))$ in quantum orthogonal
space is shown in Section 3. In Section 4, kinds of conjugations on quantum
orthogonal group and quantum orthogonal space are constructed to get quantum
AdS group and quantum AdS space both for $\vert q\vert=1$ and $q$ being real
cases. We state the differential calculus on quantum AdS space in Section 5.
Reality of differential calculus is discussed. To deduce quantum AdS algebra
from conjugation on quantum orthogonal space and quantum orthogonal group, we set up explicit
relationships between quantum group $SO_q(5)$ and quantum algebra $U_q(so(5))$
in Section 6. Section 7 is devoted to study of quantum AdS algebra.

\section{Differential calculus on quantum space}

The quantum ${\cal R}$ matrix\cite{09} for the quantum group $SO_q(5)$ is of the form

\begin{equation}
\begin{array}{rcl}
{\cal R}&=&q \displaystyle\sum_{\stackrel{i=-2}{i\not=0}}^2 e^i_i\otimes e^i_i
      + \displaystyle\sum_{\stackrel{\stackrel{i,j=-2}{i\not=j,j'}}
      {{\rm or}~i=j=0}}^2
        e^i_i\otimes e^j_j
      + \displaystyle q^{-1}\sum_{\stackrel{i=-2}{i\not=0}}^2 e^{i'}_{i'}\otimes e^i_i\\
   & &+ \displaystyle \lambda\left[\sum_{\stackrel{i,j=-2}{i>j}}^2 e^i_j\otimes e^j_i
         -\sum_{\stackrel{i,j=-2}{i>j}}^2q^{\rho_i-\rho_j}e^i_j\otimes e^{i'}_{j'}\right]~,
\end{array}
\end{equation}
where $\lambda=q-q^{-1}$, $i'=-i$ and $\rho_i=\left(\frac{3}{2},~\frac{1}{2},~0,~-\frac{1}{2},~
-\frac{3}{2}\right)$.

By making use of the ${\cal R}$ matrix, the quantum group $SO_q(5)$ with
entries
$$T=\left(\begin{array}{ccccc}
          t_{11}&t_{12}&t_{13}&t_{14}&t_{15}\\
          t_{21}&t_{22}&t_{23}&t_{24}&t_{25}\\
          t_{31}&t_{32}&t_{33}&t_{34}&t_{35}\\
          t_{41}&t_{42}&t_{43}&t_{44}&t_{45}\\
          t_{51}&t_{52}&t_{53}&t_{54}&t_{55}\end{array}\right)$$
can be written into the standard form

\begin{equation}
{\cal R}T_1T_2=T_2T_1{\cal R}~,
\end{equation}
where $T_1\equiv T\otimes 1$ and $T_2\equiv 1\otimes T$.

Of course, the $T$ matrix also should satisfy the orthogonal relations

\begin{equation}
T^tCT=C~,~~~~~~TCT^t=C~,
\end{equation}
where $C_{ij}$ is the metric on the quantum orthogonal space,
$$C=\left(\begin{array}{ccccc}
                 & & & &q^{-3/2}\\
                 & & &q^{-1/2}& \\
                 & &1& &        \\
                 &q^{1/2}& & &  \\
                q^{3/2}& & & &\end{array}\right)~.$$
The Hopf algebra structure of the quantum orthogonal group is

\begin{equation}
\begin{array}{l}
\Delta(T)=T\dot{\otimes} T~,\\
\epsilon(T)=I~,\\
S(T)=CT^tC^{-1}=\left(\begin{array}{ccccc}
                      t_{55}&q^{-1}t_{45}&q^{-3/2}t_{35}&q^{-2}t_{25}&q^{-3}t_{15}\\
                      qt_{54}&t_{44}&q^{-1/2}t_{34}&q^{-1}t_{24}&q^{-2}t_{14}\\
                      q^{3/2}t_{53}&q^{1/2}t_{43}&t_{33}&q^{-1/2}t_{23}&q^{-3/2}t_{13}\\
                      q^2t_{52}&qt_{42}&q^{1/2}t_{32}&t_{22}&q^{-1}t_{12}\\
                      q^3t_{51}&q^{2}t_{41}&q^{3/2}t_{31}&qt_{21}&t_{11}
                      \end{array}\right)~,
\end{array}
\end{equation}                      
where the operation $\dot{\otimes}$ between two matrices $A$ and $B$ is defined
as$$(A\dot{\otimes}B)_{ij}=A_{ik}\otimes B_{kj}~.$$

To define differential calculus on the non-commutative algebra generated
by the coordinates ${\bf x}(=\{x^i\},~i=-2,~-1,~0,~+1,~+2)$, we write down
the $\hat{\cal R}$ ($\equiv {\cal R}P$, and $P$ a permutation operator $P:~~~A\otimes B
=B\otimes A$) matrix explicitly,

{\small
\begin{equation}
\hat{\cal R}=\left(\begin{array}{ccccccccccccccccccccccccc}
q& & & & & & & & & & & & & & & & & & & & & & & & \\
 &\lambda& & & &1& & & & & & & & & & & & & & & & & & & \\
 & &\lambda& & & & & & & &1& & & & & & & & & & & & & & \\
 & & &\lambda& & & & & & & & & & & &1& & & & & & & & & \\
 & & & &\frac{(1-\frac{1}{q^3})}{\lambda^{-1}}& & & &-\frac{\lambda}{q^2}& &
 & &-\frac{\lambda}{q^{\frac{3}{2}}}& & & &-\frac{\lambda}{q}& & & &
 \frac{1}{q}& & & & \\
 &1& & & & & & & & & & & & & & & & & & & & & & & \\
 & & & & & &q& & & & & & & & & & & & & & & & & & \\
 & & & & & & &\lambda& & & &1& & & & & & & & & & & & & \\
 & & & &-\frac{\lambda}{q^2}& & & &\frac{1-\frac{1}{q}}{\lambda^{-1}}&
 & & &-\frac{\lambda}{q^{\frac{1}{2}}}& & & &\frac{1}{q}& & & & & & & & \\
 & & & & & & & & &\lambda& & & & & & & & & & & &1& & & \\
 & &1& & & & & & & & & & & & & & & & & & & & & & \\
 & & & & & & &1& & & & & & & & & & & & & & & & & \\
 & & & &-\frac{\lambda}{q^{\frac{3}{2}}}& & & &-\frac{\lambda}{q^{\frac{1}{2}}}&
 & & &1& & & & & & & & & & & & \\
 & & & & & & & & & & & & &\lambda& & & &1& & & & & & & \\
 & & & & & & & & & & & & & &\lambda& & & & & & & &1& & \\
 & & &1& & & & & & & & & & & & & & & & & & & & & \\
 & & & &-\frac{\lambda}{q}& & & &\frac{1}{q}& & & & & & & & & & & & & & & & \\
 & & & & & & & & & & & & &1& & & & & & & & & & & \\
 & & & & & & & & & & & & & & & & & &q& & & & & & \\
 & & & & & & & & & & & & & & & & & & &\lambda& & & &1& \\
 & & & &\frac{1}{q}& & & & & & & & & & & & & & & & & & & & \\
 & & & & & & & & &1& & & & & & & & & & & & & & & \\
 & & & & & & & & & & & & & &1& & & & & & & & & & \\
 & & & & & & & & & & & & & & & & & & &1& & & & & \\
 & & & & & & & & & & & & & & & & & & & & & & & &q
\end{array}\right)~.\end{equation}   }
The $\hat{\cal R}$ matrix is semisimple and has the spectral resolution

\begin{equation}
\begin{array}{l}
\hat{\cal R}=q{\cal P}_S-q^{-1}{\cal P}_A+q^{-4}{\cal P}_1~,\\
\hat{\cal R}^{-1}=q^{-1}{\cal P}_S-q{\cal P}_A+q^{4}{\cal P}_1~,\\
\end{array}
\end{equation}
where ${\cal P}_S$, ${\cal P}_A$ and ${\cal P}_1$ are projection operators
which act on the tensor product ${\bf x}\otimes {\bf x}$ of the fundamental
representation ${\bf x}$, and project into the symmetric, antisymmetric
and singlet irreducible representations with dimension $14$, $10$ and $1$,
respectively. It is convenient to give a concrete representation of the
projector,

$$\begin{array}{l}
{\cal P}_1=\frac{1-q^2}{(1-q^5)(1+q^3)}(C^{-1})^{ij}C_{kl}e^k_i\otimes e^l_j~,\\
{\cal P}_A=\frac{1}{q+q^{-1}}\left(-\hat{\cal R}+q-(q-q^{-4}){\cal P}_1\right)~,\\
{\cal P}_S=\frac{1}{q+q^{-1}}\left(\hat{\cal R}+q^{-1}-(q^{-1}+q^{-4}){\cal P}_1
\right)~.\end{array}$$
Analogous to the commutative case, we define the commutation relations of the
quantum orthogonal space by requiring the vanishing of their antisymmetric products.
Here the quantum antisymmetric products is given by the projector ${\cal P}_A$,

\begin{equation}
{\cal P}_A({\bf x}\otimes {\bf x})=0~.
\end{equation}
In components, the commutation relations among coordinates $x^i$ is

\begin{equation}
\begin{array}{l}
x^ix^j=qx^jx^i~,~~~~~{\rm for}~i<j~ ~{\rm and}~~i\not=-j~,\\
qx^{+2}x^{-2}-q^{-1}x^{-2}x^{+2}=\frac{q^{1/2}-q^{-1/2}}{q-1+q^{-1}}
\frac{1}{a^2}~,\\
qx^{+1}x^{-1}-q^{-1}x^{-1}x^{+1}=(1-q^2)x^{+2}x^{-2}+q\frac{q^{1/2}-
q^{-1/2}}{q-1+q^{-1}}\frac{1}{a^2}~.
\end{array}
\end{equation}
Here we have used the constraint that the quantum length ($L\propto
{\bf x}^tC{\bf x}=1/a^2$)
is invariant under quantum orthogonal group transformations.

The exterior derivative $d$\cite{10} is an operator which gives the mapping from
the coordinates to the differentials
$$d:~~~x^i\longrightarrow dx^i~.$$
It satisfies two properties
$$d^2=0~,~~~~~~~~d(fg)=(df)g+(-1)^ff(dg)~.$$
where $f$ and $g$ are $p$-forms and $(-1)^f$ is $-1(+1)$ if $f$ is odd(even)
element. The exterior derivative $d$ is invariant under the quantum group
transformation and the differential $dx^i$ transforms in the same way as the
coordinate $x^i$ under the quantum group transformation.

The derivative $\partial _i$ can be introduced by
$$d=dx^i\partial_i=C_{ij}dx^i\partial^j~.$$

For the differential calculus on the quantum orthogonal space\cite{11}, the following relations
are satisfied

\begin{equation}\label{diff}
\begin{array}{l}
x^idx^j=q\hat{R}^{ij}_{kl}dx^kx^l~,\\
{\cal P}_S(d{\bf x}\wedge d{\bf x})=0~,~~~~~
{\cal P}_1(d{\bf x}\wedge d{\bf x})=0~,\\
\partial^i x^j=(C^{-1})^{ij}+q(\hat{\cal R}^{-1})^{ij}_{kl}x^k\partial^l~,\\
\left.{\cal P}_A\right.^{ij}_{kl}\partial_j\partial_i=0~,\\
\partial^idx^j=q^{-1}\hat{\cal R}^{ij}_{kl}dx^k\partial^l~,\\
\partial^id=q^{-2}d\partial^i-(q^{-2}-q^3)\frac{1-q^2}{(1-q^5)(1+q^{-3})}
dx^iC_{jk}\partial^j\partial^k~.
\end{array}
\end{equation}

\section{Realization of quantum algebra on  quantum space}

An action of an algebra $A$ on a space $V$ is a bilinear map,

$${\cal A}:~~~A\otimes V\longrightarrow V~,~~~~p\otimes x\longrightarrow p
\cdot x~,$$
such that
$$(pq)\cdot x=p\cdot(q\cdot x)~,~~~~~1\cdot x=1~.$$
We call $V$ an $A$-module. In the same sense as comultiplication is the dual
to multiplication, coactions are dual to actions. The coaction of a coalgebra
$B$ on a space $V$ is a linear map

$$\omega_B:~~~V\longrightarrow B\otimes V~,$$
such that
$$(\omega_B\otimes {\rm id})\omega_B=({\rm id}\otimes \omega_B)\omega_B~,
~~~~~
({\rm id}\otimes\epsilon)\omega_B={\rm id}~.$$
The coalgebra is referred as a co-module.

We know that the quantum group co-acts on quantum space and is a co-module.
From the above general discussion, we conclude that, as a dual Hopf algebra
of the quantum group, the quantum algebra acts on the quantum space. Thus,
analogue to the classical case, a realization of quantum algebra on the
quantum space should be exist. And when the deformation parameter
$q\rightarrow 1$, familiar results on Lie algebra should be recovered.
Following these guidances, we present a realization of quantum algebra
$U_q(so(5))$ on the quantum orthogonal space.

For convenient, we introduce the dilatation operator $S_m$ ($m\leq 2$) as

$$\begin{array}{l}
\displaystyle S_m=1+q\lambda E_m+q^{2m+1}\lambda^2L_m\Delta_m~,\\
\displaystyle E_m=\sum_{j=-m}^mx^j\partial_j~,\\
\displaystyle\Delta_m=\sum_{j=1}^mq^{\rho_j}\partial_j\partial_{-j}+\frac{q}{1+q}
\partial_0\partial_0~,\\
\displaystyle L_m=\sum_{i=1}^m q^{\rho_i}x^{-i}x^i+\frac{q}{1+q}x^0x^0~.
\end{array}$$
The dilatation operator satisfies

\begin{equation}
S_m x^k=q^2x^k S_m~,~~~~~~~S_m\partial _k=q^{-2}\partial_kS_m~,~~~~{\rm for}~
\vert k\vert\leq m~.
\end{equation}
Using the notations

$$\begin{array}{l}
y^{-1}=x^{-1}+q^{3/2}\lambda L_1\partial_{+1}~,\\
\delta_{-1}=\partial_{-1}+q^{3/2}\lambda \Delta_1x^{+1}~,\\
y^{-2}=x^{-2}+q^{5/2}\lambda L_2\partial_{+2}~,\\
\delta_{-2}=\partial_{-2}+q^{5/2}\lambda \Delta_2x^{+2}~,
\end{array}$$
we can construct a set of independent basis on the quantum orthogonal space

\begin{equation}
\begin{array}{ll}
{\cal X}^{-2}=S_{2}^{-1/2}\mu_{+2}^{-1/2}y^{-2}~,~~~~~~&{\cal D}_{-2}=q^{-1}
S_2^{-1/2}\mu_{+2}^{-1/2}\delta_{-2}~,\\
{\cal X}^{-1}=\mu_{+2}^{-1/2}S_1^{-1/2}\mu_{+1}^{-1/2}y^{-1}~,~~~~~~&
{\cal D}_{-1}=\mu_{+2}^{-1/2}q^{-1}S_1^{-1/2}\mu_{+1}^{-1/2}\delta_{-1}~,\\
{\cal X}^0=\mu_{+2}^{-1/2}\mu_{+1}^{-1/2}x^0~,~~~~~~&{\cal D}_0=\mu_{+2}^{-1/2}
\mu_{+1}^{-1/2}\partial_0~,\\
{\cal X}^{+1}=\mu_{+2}^{-1/2}x^{+1}~,~~~~~~&{\cal D}_{+1}=\mu_{+2}^{-1/2}
\partial_{+1}~,\\
{\cal X}^2=x^2~,~~~~~~&{\cal D}_2=\partial_2~,
\end{array}
\end{equation}
where $(\mu_{\pm i})^{\pm 1}={\cal D}_{\pm i}{\cal X}^{\pm i}-
{\cal X}^{\pm i}{\cal D}_{\pm i}$ and $\mu_0^{1/2}={\cal D}_0{\cal X}^0-
{\cal X}^0{\cal D}_0$.\\
We note that the $\mu_i$'s satisfy simple commutation relations with the
independent basis ${\cal X}$ and ${\cal D}$,

$$[\mu_i,\mu_j]=0~,~~~~~~
\mu_i{\cal X}^j={\cal X}^j\mu_i\cdot\left\{\begin{array}{l}
                                           q^2~,~~~{\rm for}~i=j\\
                                           1~,~~~{\rm for}~i\not=j
                                           \end{array}\right.~,~~~~~~
\mu_i{\cal D}^j={\cal D}^j\mu_i\cdot\left\{\begin{array}{l}
                                           q^{-2}~,~~~{\rm for}~i=j\\
                                           1~,~~~{\rm for}~i\not=j
                                           \end{array}\right.~.$$

For the independent basis of coordinates and derivatives on the quantum
space\cite{0a}\cite{12}, it is not difficult to show that

\begin{equation}
\begin{array}{l}
{\cal D}_{-2}{\cal X}^{-2}=1+q^{-2}{\cal X}^{-2}{\cal D}_{-2}~,\\
{\cal D}_{-1}{\cal X}^{-1}=1+q^{-2}{\cal X}^{-1}{\cal D}_{-1}~,\\
{\cal D}_{0}{\cal X}^{0}=1+q{\cal X}^{0}{\cal D}_{0}~,\\
{\cal D}_{+1}{\cal X}^{+1}=1+q^{2}{\cal X}^{+1}{\cal D}_{+1}~,\\
{\cal D}_{+2}{\cal X}^{+2}=1+q^{2}{\cal X}^{+2}{\cal D}_{+2}~,\\[2mm]
[{\cal D}_i,{\cal D}_j]=0~,~~~~~[{\cal X}^i,{\cal X}^j]=0~,\\[2mm]
{\cal D}_i{\cal X}^j={\cal X}^j{\cal D}_i~,~~~~~{\rm for}~ i\not=j~.
\end{array}
\end{equation}

We know the Cartan matrix for its rank two Lie algebra $B_2$ is

\begin{equation}
A_{ij}=\left(\begin{array}{cc}
             2&-1\\
             -2&2\end{array}\right)~,~~~~~~\langle\alpha_i,\alpha_j\rangle
                                         =\left(\begin{array}{cc}
                                                  2&-1\\
                                                 -1&1\end{array}\right)~.
\end{equation}
So that $d_1=1$, $d_2=1/2$.

The quantum universal enveloping algebra\cite{16} $U_q\left(so(5)\right)$ is

\begin{equation}
\begin{array}{l}
[\tilde{H}_i,\tilde{H}_j]=0~,~~~~~~i,~j=1,~2,\\[1mm]
[\tilde{H}_i,\tilde{X}_j^\pm]=\pm a_{ij}\tilde{X}_j^\pm~,\\[5mm]
\displaystyle [\tilde{X}_i^+,\tilde{X}_j^-]=\delta_{ij}\frac{q_i^{\tilde{H}_i}-
q_i^{-\tilde{H}_i}}
{q_i-q_i^{-1}}~,~~~~~q_i\equiv q^{d_i}~,\\[5mm]
\displaystyle\sum_{m=0}^{1-a_{ij}}\left[\begin{array}{c}
                                        1-a_{ij}\\
                                        m\end{array}\right]_{q_i}
(\tilde{X}_i^\pm)^m \tilde{X}_j^\pm (\tilde{X}_i^\pm)^{1-a_{ij}-m}=0~,~~~~
i\not=0~,
\end{array}
\end{equation}
where $[m]_q=\frac{q^m-q^{-m}}{q-q^{-1}}$, and $\left[\begin{array}{c}
m\\
n\end{array}\right]_q=\frac{[m]_q![m-n]_q!}{[n]_q!}$.

Redefine $H_i=d_i\tilde{H}_i$, $X_i^\pm=\sqrt{[d_i]_q}\tilde{X}_i^\pm$, we have
\begin{equation}
\begin{array}{l}
[H_1,H_2]=0~,\\[1mm]
[H_1,X_1^\pm]=\pm 2X_1^\pm~,~~~~~[H_1,X_2^\pm]=\mp X_2^\pm~,\\[1mm]
[H_2,X_1^\pm]=\mp X_1^\pm~,~~~~~[H_2,X_2^\pm]=\pm X_2^\pm~,\\[2mm]
[X_1^+,X_1^-]=\displaystyle\frac{q^{H_1}-q^{-H_1}}{q-q^{-1}}~,\\[2mm]
[X_2^+,X_2^-]=\displaystyle\frac{q^{H_2}-q^{-H_2}}{q-q^{-1}}~,\\[2mm]
X_2^\pm (X_1^\pm)^2-[2]X_1^\pm X_2^\pm X_1^\pm +(X_1^\pm)^2X_2^\pm=0~,\\
X_1^\pm (X_2^\pm)^3-[3]_{\sqrt{q}}X_2^\pm X_1^\pm (X_2^\pm)^2+[3]_{\sqrt{q}}
(X_2^\pm)^2X_1^\pm X_2^\pm-(X_2^\pm)^3X_1^\pm=0~.
\end{array}
\end{equation}
The Hopf algebra structure of the quantum universal enveloping algebra
$U_q(so(5))$ is of the form

\begin{equation}
\begin{array}{l}
\Delta(H_i)=H_i\otimes 1+1\otimes H_i~,\\
\Delta(X_i^\pm)=X_i^\pm\otimes q^{H_i/2}+q^{-H_i/2}\otimes X_i^\pm~,\\
\epsilon(H_i)=0=\epsilon(X_i^\pm)~,\\
S(H_i)=-H_i~,~~~~~~S(X_i^\pm)=-q^{\mp d_i}X_i^\pm~.
\end{array}
\end{equation}

In terms of the set of independent basis ${\cal X}^i,~{\cal D}_i$ on the
quantum space, we write down  a realization
of the quantum universal enveloping algebra $U_q(so(5))$ in fundamental
representation,

\begin{equation}
\begin{array}{l}
q^{H_1}=\left(\frac{\mu_{-2}\mu_{+1}}{\mu_{-1}\mu_{+2}}
\right)^{1/2}~,\\
X^-_1=\sqrt{q^{-1}\lambda}\left(\frac{\mu_{-1}\mu_{+2}}{\mu_{-2}\mu_{+1}}\right)^{1/4}
\mu_{+2}^{-1/2}
\left((\mu_{-2}\mu_{+1})^{1/2}{\cal X}^{-1}
{\cal D}_{-2}-q{\cal X}^{+2}{\cal D}_{+1}\right)~,\\
X^+_1=\sqrt{q^{-1}\lambda}\left(\frac{\mu_{-1}\mu_{+2}}{\mu_{-2}\mu_{+1}}\right)^{1/4}
\mu_{+2}^{-1/2}
\left((\mu_{-2}\mu_{+1})^{1/2}{\cal X}^{-2}
{\cal D}_{-1}-q{\cal X}^{+1}{\cal D}_{+2}\right)~,\\
q^{H_2}=\left(\frac{\mu_{-1}}{\mu_{+1}}\right)^{1/2}~,\\
X^-_2=\sqrt{q^{1/2}\lambda}(\mu_{-1}\mu_{+1})^{-1/4}\left(q^{-3/2}\mu_{-1}^{1/2}
{\cal X}^0{\cal D}_{-1}-{\cal X}^{+1}{\cal D}_0\right)~,\\
X^+_2=\sqrt{q^{1/2}\lambda}(\mu_{-1}\mu_{+1})^{-1/4}\left(q^{1/2}\mu_{-1}^{1/2}
{\cal X}^{-1}{\cal D}_0-{\cal X}^0{\cal D}_{+1}\right)~.
\end{array}
\end{equation}
Therefore, we indeed have a natural action of the quantum algebra $U_q(so(5))$
on the
quantum orthogonal space. It may be referred as the quantum counterpart of
the Lie algebra realized on Euclidean space. In fact, at the case of
$q=1$, this realization recovered a ordinary realization of Lie algebra.
This is helpful for clearly understanding of meanings of quantum group and quantum
algebra.
\section{Quantum AdS group and quantum AdS space}

To study quantum AdS group, real forms and $*$-conjugations of quantum
orthogonal group should be introduced. A $*$-structure on a Hopf algebra $A$
is an algebra anti-automorphism $(\eta ab)^*=\bar{\eta}b^*a^*~(\forall a,~b\in
 A,~\forall\eta\in C$), coalgebra automorphism $\Delta\cdot *=(*\otimes *)
 \cdot\Delta$, $\epsilon\cdot *=\epsilon$ and involution $*^2=1$.

Following FRT\cite{09}, on quantum orthogonal groups conjugations can be defined.
The first type is trivially as $T^\times =T$.
Compatibility with the Yang-Baxter
equation requires $\overline{\cal R}={\cal R}^{-1}$ and
$\overline{C}C^t=1$. This occurs only for
$\vert q\vert=1$. Then the $CTT$ relations are invariant under the
$*$-conjugation.  As we know the quantum orthogonal group
co-acts on the quantum orthogonal space, and may induce an associated
conjugation on the quantum space as well. More precisely a conjugation
on the quantum space is compatible with a conjugation on its  quantum symmetry
group if the coaction $\omega(x^a)=T^a_{~b}\otimes x^b$ satisfies
$\omega({\bf x}^\times)=T^\times\otimes {\bf x}^\times\equiv \delta^\times
({\bf x})$. The unique associated
quantum space conjugation here is $(x^a)^\times=x^a$.
Clearly, we can not get the desired quantum AdS space by this kind of
conjugation on quantum orthogonal space.
Further structure should be added
to the conjugation to arrive at the quantum AdS group and quantum AdS space.
For the aim, we introduce another operation on the quantum orthogonal group as

\begin{equation}
T^\dagger={\cal D}T{\cal D}^{-1}~,
\end{equation}
where the matrix ${\cal D}$ is of the form
$${\cal D}=\left(\begin{array}{ccccc}
                 1& & & & \\
                  &1& & & \\
                  & &-1&& \\
                  & & &1& \\
                  & & & &1\end{array}\right)~.$$
It is easy to check that the ${\cal D}$ matrix is a special element of the
quantum orthogonal group\cite{13}, {\it i.e.},
$$\begin{array}{l}
{\cal R}_{12}{\cal D}_1{\cal D}_2={\cal D}_2{\cal D}_1{\cal R}_{12}~,\\
{\cal D}^tC{\cal D}=C~,~~~~{\cal D}C{\cal D}^t=C~.
\end{array}$$
Thus the $\dagger$ operation is compatible with the Hopf algebra structure
of the quantum orthogonal group,
$$
\begin{array}{l}
{\cal R}_{12}T^\dagger_1T^\dagger_2=T^\dagger_2T^\dagger_1{\cal R}_{12}~,\\
(T^\dagger)^tCT^\dagger=C~,~~~~~T^\dagger C(T^\dagger)^t=C~,\\
\Delta(T^\dagger)={\cal D}\Delta(T){\cal D}^{-1}=T^\dagger\otimes
T^\dagger~,\\
\epsilon(T^\dagger)=\epsilon(T)~,~~~~~S(T^\dagger)=[S(T)]^\dagger~.
\end{array}$$
We finally obtain the desired AdS quantum group by using the combined operation of the
$\times$ and $\dagger$, i.e., the conjugation $T^*\equiv
T^{\times\dagger}={\cal D}T{\cal D}^{-1}$.
The induced conjugation on the quantum space is ${\bf x}^*\equiv {\bf x}^{
\times\dagger}={\cal D}{\bf x}$.
To show that the conjugation really gives the quantum AdS group and quantum
AdS space, we should find a linear transformation ${\bf x}\longrightarrow
{\bf x}'=M {\bf x},~ T\longrightarrow T'=MTM^{-1}$
such that the new coordinates ${\bf x}'$ and $T'$ are real and the new metric
$C'=(M^{-1})^tCM^{-1}$ diagonal in the $q\longrightarrow 1$ limit,
$C'\vert_{q=1}={\rm diag}(1,-1,-1,-1,1)$.
Here the $M$ matrix is given by
$$M=\frac{1}{\sqrt{2}}\left(\begin{array}{ccccc}
                            1&  &  &  &1\\
                           -1&  &  &  &1\\
                             &  &i\sqrt{2}& &\\
                             &-1&  & 1&  \\
                             & 1&  & 1& \end{array}\right)~.$$

The second conjugation of the quantum orthogonal group given by FRT\cite{09}
is realized via the
metric, {\it i.e.},  $T^\star=C^tTC^t$. The condition on the quantum
${\cal R}$ matrix is
$\overline{\cal R}={\cal R}$, which happens for $q\in R$. Again the $CTT$
relations are invariant under such a conjugation operation.
The corresponding real form is $SO_q(5;R)$.
The induced conjugation on the quantum space is ${\bf x}^\star=C^t{\bf x}$.
Again, we can not get the desired quantum AdS group and quantum AdS space by
this kind of conjugation along.
To obtain quantum AdS group, we introduce the operation $\ddagger$ on the
quantum orthogonal group $SO_q(5)$ as

\begin{equation}
T^\ddagger={\cal N}T{\cal N}^{-1}~,
\end{equation}
where ${\cal N}$ matrix is of the form
$${\cal N}=\left(\begin{array}{ccccc}
                  1&  &  &  &  \\
                   &-1&  &  &  \\
                   &  &-1&  &  \\
                   &  &  &-1&  \\
                   &  &  &  &1\end{array}\right)~.$$
The ${\cal N}$ matrix is also a special element of the
quantum orthogonal group, {\it i.e.},

\begin{equation}\label{rnn}
\begin{array}{l}
{\cal R}_{12}{\cal N}_1{\cal N}_2={\cal N}_2{\cal N}_1{\cal R}_{12}~,\\
{\cal N}^tC{\cal N}=C~,~~~~{\cal N}C{\cal N}^t=C~.
\end{array}
\end{equation}
Thus the $\ddagger$ operation is also compatible with the Hopf algebra
structure of quantum group.
We obtain the desired AdS quantum group by using the combined operation of the
$\star$ and $\ddagger$, {\it i.e.}, the conjugation $T^*\equiv
T^{\star\ddagger}={\cal N}C^tTC^t{\cal N}^{-1}$.
The induced conjugation on the quantum space is ${\bf x}^*\equiv
{\bf x}^{\star\ddagger}=C^t {\cal N}{\bf x}$.

Parallel to the case of $\vert q\vert=1$, we prove the combined conjugation really
gives the quantum AdS group and quantum AdS space by trying to
find a linear transformation ${\bf x}\longrightarrow {\bf x}'=N {\bf x},~ T\longrightarrow
T'=NTN^{-1}$
such that the new coordinates ${\bf x}'$ and $T'$ are real and the new metric
$C'=(N^{-1})^tCN^{-1}$ diagonal in the $q\longrightarrow 1$ limit,
$C'\vert_{q=1}={\rm diag}(1,-1,-1,-1,1)$. We find a $N$ matrix which
satisfies all these requirements of the form
$$N=\frac{1}{\sqrt{2}}\left(\begin{array}{ccccc}
                            1&  &  &  &q^{3/2}\\
                             & -i&  &-iq^{1/2}&  \\
                             &  &i\sqrt{2}& &\\
                             &q^{-1/2}&  &-1&  \\
                    iq^{-3/2}&  &  &  &-i \end{array}\right)~.$$

\section{Differential Calculus on the quantum AdS space}

To study differential calculus on the quantum AdS space, we should
introduce conjugate derivatives. Because there are two different conjugations
on the quantum AdS space corresponds to the deformation parameter $q$ being root of unit
or real, respectively. We present differential calculus on the quantum AdS
space for the two cases separately.

In the case of $q$ being root of unit, we note the conjugation on $x^i$,
$\hat{\cal R}$ and $C$ is given by

\begin{equation}
\left.x^i\right.^*={\cal D}_{ij}x^j~,~~~~~\overline{\hat{\cal R}^{uv}_{ij}}=\left.
\hat{\cal R}^{-1}\right.
^{vu}_{ji}~,~~~~~\overline{C^{ij}}=C^{ji}~.
\end{equation}
Because the matrix ${\cal D}$ is a special element of the quantum orthogonal group,
we can prove that

$$\left.\hat{\cal R}^{-1}\right.^{ik}_{ld}
{\cal D}_{kn}{\cal D}_{sl}{\cal D}_{ip}{\cal D}_{td}=
\left.\hat{\cal R}^{-1}\right.^{st}_{pn}~.$$
The $\hat{\cal R}$-matrix has the following symmetry properties

\begin{equation}\label{I}
C^{im}\hat{\cal R}^{jn}_{mk}C_{nl}=\left.\hat{\cal R}^{-1}\right._{kl}^{ij}
=C_{km}\hat{\cal R}^{mi}_{ln}C^{nj}~,
\end{equation}
and

\begin{equation}\label{II}
\hat{\cal R}^{ij}_{kl}=\hat{\cal R}^{kl}_{ij}~.
\end{equation}
Using the above relations as well as  the conjugation of the definition of
derivatives,
$\partial_ix^j=\delta^j_i+q\hat{\cal R}^{jk}_{il}x^l\partial_k$, we obtain the
action of the conjugate derivatives on the quantum AdS space,

\begin{equation}
\hat{\partial}_mx^s=\delta^s_m+q\hat{\cal R}^{sl}_{mk}x^k\hat{\partial}_l~,
\end{equation}
where $\hat{\partial}_m\equiv -q^{-1}{\cal D}_{mk}C_{kv}\left.\partial^v\right.^*$.

This show that the action of the conjugate derivatives on the quantum AdS space
is almost the same with that of the derivatives at the case of $\vert q\vert=1$.
In fact, we can represent
the conjugate derivatives $\partial^*$ linearly in terms of the
derivatives $\partial$,

\begin{equation}
\left.\partial_v\right.^*=-qC_{vk}{\cal D}_{km}\partial_m~.
\end{equation}

From a very general consideration\cite{10}, we know that there are two types of
consistent derivatives $\partial_i$ and $\tilde{\partial}_i$ satisfy the
following relations, respectively,

\begin{equation}\label{conjugate}
\begin{array}{l}
\partial_ix^j=\delta^j_i+q\hat{\cal R}^{jk}_{il}x^l\partial_k~,\\
\tilde{\partial}_kx^v=\delta^v_k+q^{-1}\left.\hat{\cal R}^{-1}\right.^{vi}_{kj}
x^j\tilde{\partial}_i~.
\end{array}
\end{equation}
The first one is just what given in Eq.(\ref{diff}). We show that the second
one is related with the conjugate derivatives in the case of $q\in R$.

In the following, we present actions of the conjugate derivatives on the quantum
AdS space at the case of $q\in R$.
To proceed, we first conjugate the first relation in Eq.(\ref{conjugate})
and invert it to find an
expression for $\partial_i^*x^j$ in terms of $x^j\partial_i^*$ by using
$$\left.x^i\right.^*=C_{ji}{\cal N}_{jk}x^k$$
and the symmetry properties of the $\hat{\cal R}$-matrix Eq.(\ref{I}) and (\ref{II}).
The result is:

\begin{equation}
-q^5{\cal N}_{ip}C_{ia}C^{va}\left.\partial^v\right.^*x^s=q^{4}C_{ia}C_{p\delta}
\hat{\cal R}^{p\delta}_{al}{\cal N}_{sl}{\cal N}_{ip}-q^4\left.\hat{\cal R}^{-1}\right.^{ik}_{ld}
{\cal N}_{kn}{\cal N}_{sl}{\cal N}_{ip}{\cal N}_{td}x^nC_{db}C^{mb}
{\cal N}_{dt}\left.\partial^m\right.^*~.
\end{equation}
Making use of Eq.(\ref{rnn}), we know that

$$\left.\hat{\cal R}^{-1}\right.^{ik}_{ld}
{\cal N}_{kn}{\cal N}_{sl}{\cal N}_{ip}{\cal N}_{td}=
\left.\hat{\cal R}^{-1}\right.^{st}_{pn}~.$$
Then, we get

\begin{equation}
\hat{\partial}_px^s=\delta^s_p+q^{-1}\left.\hat{\cal R}^{-1}\right.^{st}_{pn}
x^n\hat{\partial}_t~,
\end{equation}
where
$$\hat{\partial}_p=-q^5{\cal N}_{ip}C_{ia}C^{va}\left.\partial^v\right.^*~.$$

$\hat{\partial}_i$ (or equivalient $\partial_i^*$)
can be expressed algebraically in terms of $x^k$ and $\partial_l$\cite{14},

\begin{equation}
\hat{\partial}_k=q^3S_2^{-1}[\Delta_2,x^k]~.
\end{equation}
To complete the differential calculus on the quantum AdS space, we give the
consistent relations satisfied by $\hat{\partial}$ and $\partial$

\begin{equation}
\begin{array}{l}
\hat{\partial}_i\hat{\partial}_j=q^{-1}\hat{\cal R}^{lk}_{ji}\hat{\partial}_k
\hat{\partial}_l~,\\
\hat{\partial}_i\partial_j=q\hat{\cal R}^{lk}_{ji}\partial_k
\hat{\partial}_l~.
\end{array}
\end{equation}

\section{Quantum group via quantum algebra}

It is well known that, in Lie group theory, there is an exponential
corresponding between group and algebra. Representations of Lie group are
easily deduced from that of Lie algebra. And then quantum field theory
is constructed based on representations of Lie algebra. However, in general,
there is no such a direct transformation from quantum algebra to quantum group,
besides duality between them. Bearing the goal of constructing quantum field
theory on noncommutative geometry, we try to set up an explicit relationship\cite{15}
between quantum orthogonal group and quantum orthogonal algebra.

It is convenient to introduce the operator $L^+$ as

\begin{equation}
L^+=\left(\begin{array}{ccccc}
          l^+_{11}&l^+_{12}&l^+_{13}&l^+_{14}&l^+_{15}\\
                  &l^+_{22}&l^+_{23}&l^+_{24}&l^+_{25}\\
                  &        &l^+_{33}&l^+_{34}&l^+_{35}\\
                  &        &        &l^+_{44}&l^+_{45}\\
                  &        &        &        &l^+_{55}\end{array}\right)~,
\end{equation}
where
\begin{equation}
\begin{array}{l}
l^+_{11}=q^{H_1+H_2}~,\\
l^+_{12}=\lambda q^{-1/2}X_1^+q^{\frac{H_1}{2}+H_2}~,\\
l^+_{13}=\lambda q^{-1/2}\left(X_1^+X_2^+
         -q^{-1}X_2^+X_1^+\right)q^{\frac{H_1+H_2}{2}}~,\\
l^+_{14}=\lambda q^{-5/2}\left(-qX_1^+(X_2^+)^2+(q+1)
         X_2^+X_1^+X_2^+-(X_2^+)^2X_1^+\right)
         q^{H_1/2}~,\\
l^+_{15}=\frac{\lambda^2 q^{-2}}{1+q}\left((X_1^+)^2(X_2^+)^2-
(q+1+q^{-1})X_1^+X_2^+X_1^+X_2^++(q+1+q^{-1})X_1^+(X_2^+)^2X_1^+\right.\\
~~~~~~~~~~~~~~~~\left. -(q+1+q^{-1})X_2^+X_1^+X_2^+X_1^++X_2^+(X_1^+)^2X_2^+
         +(X_2^+)^2(X_1^+)^2\right)~,\\
l^+_{22}=q^{H_2}~,\\
l^+_{23}=\lambda q^{-1/2}X_2^+q^{H_2/2}~,\\
l^+_{24}=-\frac{\lambda^2}{q(1+q)}(X_2^+)^2~,\\
l^+_{25}=\lambda q^{-5/2}\left(X_1^+(X_2^+)^2-(1+q)
         X_2^+X_1^+X_2^++q(X_2^+)^2X_1^+\right)q^{-H_1/2}~,\\
l^+_{33}=1~,\\
l^+_{34}=-\lambda q^{-1}X_2^+q^{-H_2/2}~,\\
l^+_{35}=\lambda q^{-1}\left(-q^{-1}X_1^+X_2^++
         X_2^+X_1^+\right)q^{-\frac{H_1+H_2}{2}}~,\\
l^+_{44}=q^{-H_2}~,\\
l^+_{45}=-\lambda q^{-1/2}X_1^+q^{-(\frac{H_1}{2}+H_2)}~,\\
l^+_{55}=q^{-(H_1+H_2)}~.
\end{array}
\end{equation}
In the same manner, we introduce another operator $L^-$ as

\begin{equation}
L^-=\left(\begin{array}{ccccc}
          l^-_{11}&        &        &        &        \\
          l^-_{21}&l^-_{22}&        &        &        \\
          l^-_{31}&l^-_{32}&l^-_{33}&        &        \\
          l^-_{41}&l^-_{42}&l^-_{43}&l^-_{44}&        \\
          l^-_{51}&l^-_{52}&l^-_{53}&l^-_{54}&l^-_{55}\end{array}\right)~,
\end{equation}
where
\begin{equation}
\begin{array}{l}
l^-_{11}=q^{-(H_1+H_2)}~,\\
l^-_{21}=-\lambda q^{1/2}X_1^-q^{-\left(\frac{H_1}{2}+H_2\right)}~,\\
l^-_{31}=\lambda  q^{3/2}\left(X_1^-X_2^-
         -q^{-1}X_2^-X_1^-\right)q^{-\frac{H_1+H_2}{2}}~,\\
l^-_{41}=\lambda q^{1/2}\left(qX_1^-(X_2^-)^2-(q+1)
         X_2^-X_1^-X_2^-+(X_2^-)^2X_1^-\right)q^{-H_1/2}~,\\
l^-_{51}=\frac{\lambda^2 q^{2}}{1+q}\left((X_1^-)^2(X_2^-)^2-
(q+1+q^{-1})X_1^-X_2^-X_1^-X_2^-+(q+1+q^{-1})X_1^-(X_2^-)^2X_1^-\right.\\
~~~~~~~~~~~~~~~~\left. -(q+1+q^{-1})X_2^-X_1^-X_2^-X_1^-+X_2^-(X_1^-)^2X_2^-
         +(X_2^-)^2(X_1^-)^2\right)~,\\
l^-_{22}=q^{-H_2}~,\\
l^-_{32}=-\lambda q^{1/2}X_2^-q^{-H_2/2}~,\\
l^-_{42}=-\frac{\lambda^2 q}{1+q}(X_2^-)^2~,\\
l^-_{52}=\lambda q^{1/2}\left(-X_1^-(X_2^-)^2+(1+q)
         X_2^-X_1^-X_2^- -q(X_2^-)^2X_1^-\right)q^{H_1/2}~,\\
l^-_{33}=1~,\\
l^-_{43}=\lambda X_2^-q^{H_2/2}~,\\
l^-_{53}=\lambda q\left(-q^{-1}X_1^-X_2^-+X_2^-X_1^-\right)q^{\frac{H_1+H_2}{2}}~,\\
l^-_{44}=q^{H_2}~,\\
l^-_{54}=\lambda q^{1/2}X_1^-q^{\frac{H_1}{2}+H_2}~,\\
l^-_{55}=q^{H_1+H_2}~.
\end{array}
\end{equation}
By making use of the operators $L^\pm$, we can write the quantum universal
enveloping algebra $U_q(so(5))$ into a compact form

\begin{equation}\label{algebra}
\begin{array}{l}
{\cal R}_{12}L^\pm_1L^\pm_2=L^\pm_2 L^\pm_1{\cal R}_{12}~,\\
{\cal R}_{12}L^-_1L^+_2=L^+_2 L^-_1{\cal R}_{12}~,\\
\Delta(L^\pm)=L^\pm\dot{\otimes}L^\pm~,~~~~~~\epsilon(L^\pm)=1~,\\
S(L^+)=\left(\begin{array}{ccccc}
             l^+_{55}&q^{-1}l^+_{45}&q^{-3/2}l^+_{35}&q^{-2}l^+_{25}&
             q^{-3}l^+_{15}\\
             &l^+_{44}&q^{-1/2}l^+_{34}&q^{-1}l^+_{24}&q^{-2}l^+_{14}\\
             & &l^+_{33}&q^{-1/2}l^+_{23}&q^{-3/2}l^+_{13}\\
             & & &l^+_{22}&q^{-1}l^+_{21}\\
             & & & &l^+_{11}\end{array}\right)~,\\
S(L^-)=\left(\begin{array}{ccccc}
             l^-_{55}& & & &\\
             ql^-_{54}&l^-_{44}& & &\\
             q^{3/2}l^-_{53}&q^{1/2}l^-_{43}&l^-_{33}& &\\
             q^2l^-_{52}&ql^-_{42}&q^{1/2}l^-_{32}&l^-_{22}&\\
             q^{3}l^-_{51}&q^{2}l^-_{41}&q^{3/2}l^-_{31}&ql^-_{21}&l^-_{11}
             \end{array}\right)~.
\end{array}
\end{equation}

As well known, in the classical case, many important properties of Lie group
are conveniently studied using corresponding Lie algebra and vise verse. This
is based on the exponential relation of Lie group and Lie algebra. However,
in the quantum group theory, there is no such a counterpart of exponential
has been found. All we know is the duality between quantum group and quantum
algebra.
We give here an explicit relation between elements of quantum group and
generators
of the quantum universal enveloping algebra,

\begin{equation}
T=L^+\dot{\otimes} L^-~,~~~~~{\rm or}~~~T=L^-\dot{\otimes} L^+~.
\end{equation}
Then elements of the $T$ matrix are expressed as

\begin{equation}
\begin{array}{l}
t_{11}=l_{11}^+\otimes l_{11}^-+l_{12}^+\otimes l_{21}^-+l_{13}^+\otimes l_{31}^-
        +l_{14}^+\otimes l_{41}^-+l_{15}^+\otimes l_{51}^-~,\\
t_{12}=l_{12}^+\otimes l_{22}^-+l_{13}^+\otimes l_{32}^-+l_{14}^+\otimes l_{42}^-
        +l_{15}^+\otimes l_{52}^-~,\\
t_{13}=l_{13}^+\otimes l_{33}^-+l_{14}^+\otimes l_{43}^-+l_{15}^+\otimes l_{53}^-~,\\
t_{14}=l_{14}^+\otimes l_{44}^-+l_{15}^+\otimes l_{54}^-~,\\
t_{15}=l_{15}^+\otimes l_{55}^-~,\\
t_{21}=l_{22}^+\otimes l_{21}^-+l_{23}^+\otimes l_{31}^-+l_{24}^+\otimes l_{41}^-
        +l_{25}^+\otimes l_{51}^-~,\\
t_{22}=l_{22}^+\otimes l_{22}^-+l_{23}^+\otimes l_{32}^-+l_{24}^+\otimes l_{42}^-
        +l_{25}^+\otimes l_{52}^-~,\\
t_{23}=l_{23}^+\otimes l_{33}^-+l_{24}^+\otimes l_{43}^-+l_{25}^+\otimes l_{53}^-~,\\
t_{24}=l_{24}^+\otimes l_{44}^-+l_{25}^+\otimes l_{54}^-~,\\
t_{25}=l_{25}^+\otimes l_{55}^-~,\\
t_{31}=l_{33}^+\otimes l_{31}^-+l_{34}^+\otimes l_{41}^-+l_{35}^+\otimes l_{51}^-~,\\
t_{32}=l_{33}^+\otimes l_{32}^-+l_{34}^+\otimes l_{42}^-+l_{35}^+\otimes l_{52}^-~,\\
t_{33}=l_{33}^+\otimes l_{33}^-+l_{34}^+\otimes l_{43}^-+l_{35}^+\otimes l_{53}^-~,\\
t_{34}=l_{34}^+\otimes l_{44}^-+l_{35}^+\otimes l_{54}^-~,\\
t_{35}=l_{35}^+\otimes l_{55}^-~,\\
t_{41}=l_{44}^+\otimes l_{41}^-+l_{45}^+\otimes l_{51}^-~,\\
t_{42}=l_{44}^+\otimes l_{42}^-+l_{45}^+\otimes l_{52}^-~,\\
t_{43}=l_{44}^+\otimes l_{43}^-+l_{45}^+\otimes l_{53}^-~,\\
t_{44}=l_{44}^+\otimes l_{44}^-+l_{45}^+\otimes l_{54}^-~,\\
t_{45}=l_{45}^+\otimes l_{55}^-~,\\
t_{51}=l_{55}^+\otimes l_{51}^-~,\\
t_{52}=l_{55}^+\otimes l_{52}^-~,\\
t_{53}=l_{55}^+\otimes l_{53}^-~,\\
t_{54}=l_{55}^+\otimes l_{54}^-~,\\
t_{55}=l_{55}^+\otimes l_{55}^-~;
\end{array}
\end{equation}     
or, equivalently

\begin{equation}
\begin{array}{l}
t_{11}=l_{11}^-\otimes l_{11}^+~,\\
t_{12}=l_{11}^-\otimes l_{12}^+~,\\
t_{13}=l_{11}^-\otimes l_{13}^+~,\\
t_{14}=l_{11}^-\otimes l_{14}^+~,\\
t_{15}=l_{11}^-\otimes l_{15}^+~,\\
t_{21}=l_{21}^-\otimes l_{11}^+~,\\
t_{22}=l_{21}^-\otimes l_{12}^++l_{22}^-\otimes l_{22}^+~,\\
t_{23}=l_{21}^-\otimes l_{13}^++l_{22}^-\otimes l_{23}^+~,\\
t_{24}=l_{21}^-\otimes l_{14}^++l_{22}^-\otimes l_{24}^+~,\\
t_{25}=l_{21}^-\otimes l_{15}^++l_{22}^-\otimes l_{25}^+~,\\
t_{31}=l_{31}^-\otimes l_{11}^+~,\\
t_{32}=l_{31}^-\otimes l_{12}^++l_{32}^-\otimes l_{22}^+~,\\
t_{33}=l_{31}^-\otimes l_{13}^++l_{32}^-\otimes l_{23}^++l_{33}^-\otimes l_{33}^+~,\\
t_{34}=l_{31}^-\otimes l_{14}^++l_{32}^-\otimes l_{24}^++l_{33}^-\otimes l_{34}^+~,\\
t_{35}=l_{31}^-\otimes l_{15}^++l_{32}^-\otimes l_{25}^++l_{33}^-\otimes l_{35}^+~,\\
t_{41}=l_{41}^-\otimes l_{11}^+~,\\
t_{42}=l_{41}^-\otimes l_{12}^++l_{42}^-\otimes l_{22}^+~,\\
t_{43}=l_{41}^-\otimes l_{13}^++l_{42}^-\otimes l_{23}^++l_{43}^-\otimes l_{33}^+~,\\
t_{44}=l_{41}^-\otimes l_{14}^++l_{42}^-\otimes l_{24}^++l_{43}^-\otimes l_{34}^+
       +l_{44}^-\otimes l_{44}^+~,\\
t_{45}=l_{41}^-\otimes l_{15}^++l_{42}^-\otimes l_{25}^++l_{43}^-\otimes l_{35}^+
       +l_{44}^-\otimes l_{45}^+~,\\
t_{51}=l_{51}^-\otimes l_{11}^+~,\\
t_{52}=l_{51}^-\otimes l_{12}^++l_{52}^-\otimes l_{22}^+~,\\
t_{53}=l_{51}^-\otimes l_{13}^++l_{52}^-\otimes l_{23}^++l_{53}^-\otimes l_{33}^+~,\\
t_{54}=l_{51}^-\otimes l_{14}^++l_{52}^-\otimes l_{24}^++l_{53}^-\otimes l_{34}^+
       +l_{54}^-\otimes l_{44}^+~,\\
t_{55}=l_{51}^-\otimes l_{15}^++l_{52}^-\otimes l_{25}^++l_{53}^-\otimes l_{35}^+
       +l_{54}^-\otimes l_{45}^++l_{55}^-\otimes l_{55}^+~.
\end{array}
\end{equation}
It is straightforward to verify that the matrix $T$ defined above really
satisfies all relations which a quantum group should be satisfied.
Let $P^\otimes$ be the transposition operator, for arbitrary matrices $A$ and
$B$, which satisfies
$$P^\otimes:~~~A\dot{\otimes}B=B\dot{\otimes}A~.$$
Then the comultiplication $\Delta$ for these tensor operator $T$ can be
introduced as

\begin{equation}
\Delta=({\rm id}\otimes P^\otimes \otimes{\rm id})\Delta_{L^\pm}\dot{\otimes}
\Delta_{L^\mp}~.
\end{equation}
It is easy to check that

$$\begin{array}{rcl}
  \Delta (T)&=&({\rm id}\otimes P^\otimes \otimes{\rm id})\Delta_{L^\pm}\dot{\otimes}
               \Delta_{L^\mp}(L^\pm\dot{\otimes}L^\mp)\\
            &=&({\rm id}\otimes P^\otimes \otimes{\rm id})
               (L^\pm\dot{\otimes}L^\pm\dot{\otimes}L^\mp\dot{\otimes}L^\mp)\\
            &=&(L^\pm\dot{\otimes}L^\mp)\dot{\otimes}(L^\pm\dot{\otimes}
               L^\mp)\\
            &=&T\dot{\otimes}T~.\end{array}$$
Define counit operator $\epsilon$ as

$$\epsilon=\epsilon_{L^\pm}\otimes\epsilon_{L^\mp}~.$$
Then we have $\epsilon(T)=1$. Finally, the antipode operator $S$ is of the
form,
$$S=P(S_{L^\mp}\otimes S_{L^\pm})P^\otimes~.$$
By making use of the relations (\ref{algebra}), it is a straightforward calculation to check that

$$\begin{array}{rcl}
S(T)&=&S(L^\pm\dot{\otimes}L^\mp)\\
    &=&P(S_{L^\mp}\otimes S_{L^\pm})P^\otimes(L^\pm\dot{\otimes}L^\mp)\\
    &=&P(S_{L^\mp}(L^\mp)\otimes S_{L^\pm}(L^\pm))\\
    &=&CT^tC^{-1}~.\end{array}$$
At the stage, we can say that a realization of quantum group $SO_q(5)$ is
given in terms of quantum algebra $U_q(so(5))$.

\section{Quantum AdS Algebra}

Quantum field theory is usually constructed based on representations of an algebra. To
get a quantum field theory on noncommutative geometry, we should investigate properties of the quantum
AdS algebra. Of course, first of all, we should give conjugations on quantum
AdS algebra, which is consistent with conjugations on quantum AdS group and quantum
AdS space.

At the case of $\vert q\vert=1$, conjugation on quantum AdS group is given by

\begin{equation}
\begin{array}{rcl}
T^*&=&{\cal D}T{\cal D}^{-1}\\
&=&\left(\begin{array}{ccccc}
          t_{11}&t_{12}&-t_{13}&t_{14}&t_{15}\\
          t_{21}&t_{22}&-t_{23}&t_{24}&t_{25}\\
          -t_{31}&-t_{32}&t_{33}&-t_{34}&-t_{35}\\
          t_{41}&t_{42}&-t_{43}&t_{44}&t_{45}\\
          t_{51}&t_{52}&-t_{53}&t_{54}&t_{55}\end{array}\right)~.
\end{array}
\end{equation}
From the relationship between quantum group and quantum algebra, {\it i.e.},
$T=L^\pm\dot{\otimes} L^\mp$, we deduce directly the corresponding quantum AdS algebra  as

\begin{equation}
\begin{array}{l}
\left.X_1^+\right.^*=X_1^+~,~~~~~
\left.X_1^-\right.^*=X_1^-~,~~~~~
H_1^*=-H_1~,\\
\left.X_2^+\right.^*=-X_2^+~,~~~~~
\left.X_2^-\right.^*=-X_2^-~,~~~~~
H_2^*=-H_2~.
\end{array}
\end{equation}
It is easy to show that this type of conjugation on quantum algebra satisfies
all requirements for a conjugation.

For the case of $q$ being real, we know from the previous section
that the conjugation structure of the quantum AdS group is given by

\begin{equation}\label{mmm}
\begin{array}{rcl}
T^*&=&{\cal N}C^tT(C^{-1})^t{\cal N}^{-1}\\
             &=&\left(\begin{array}{ccccc}
                      t_{55}&-qt_{54}&-q^{3/2}t_{53}&-q^2t_{52}&q^3t_{51}\\
                      -q^{-1}t_{45}&t_{44}&q^{1/2}t_{43}&qt_{42}&-q^2t_{41}\\
                      -q^{-3/2}t_{35}&q^{-1/2}t_{34}&t_{33}&q^{1/2}t_{32}&
                      -q^{3/2}t_{31}\\
                      -q^{-2}t_{25}&q^{-1}t_{24}&q^{-1/2}t_{23}&t_{22}&
                      -qt_{21}\\
                      q^{-3}t_{15}&-q^{-2}t_{14}&-q^{-3/2}t_{13}&-q^{-1}t_{12}&
                      t_{11}\end{array}\right)~.
\end{array}
\end{equation}
For $T=L^\pm\dot{\otimes}L^\mp$, the proper conjugation operation is given by

\begin{equation}\label{nnn}
T^*=\left.L^\mp\right.^*\dot{\otimes}\left.L^\pm\right.^*~,~~~~~~
\left(\left.L^\pm\right.^*\right)_{ij}\equiv \left(L^\pm_{ij}\right)^*~.
\end{equation}
Comparing of Eq.(\ref{mmm}) and (\ref{nnn}) gives the corresponding quantum
AdS algebra to quantum AdS group and quantum AdS space,

\begin{equation}
\begin{array}{l}
\left.X_1^+\right.^*=-q^2X_1^-~,~~~~~
\left.X_1^-\right.^*=-q^{-2}X_1^+~,~~~~~
H_1^*=H_1~,\\
\left.X_2^+\right.^*=q^{3/2}X_2^-~,~~~~~
\left.X_2^-\right.^*=q^{-3/2}X_2^+~,~~~~~
H_2^*=H_2~.
\end{array}
\end{equation}
Therefore, finally, we obtain the quantum AdS algebra for both the case of
$\vert q\vert=1$ and $q\in R$. The representations of the quantum AdS
algebra\cite{17} should be foundations of constructing quantum field theory on the
quantum AdS space. Further investigating on this direction is still on
progressing.

\bigskip
\bigskip
\centerline{\large\bf Acknowledgments }

I would like to thank Prof. J. Wess for introducing the problem to me and for
enlightening discussions.
I am grateful to B. L. Cerchiai and H. Steinacker for valuable discussions.
The work was supported in part by the National Science Foundation of
China under Grant 19625512.

\end{document}